# Cross-Subreddit Behavior as Open-Source Indicators of Coordinated Influence: A Case Study of r/Sino & r/China

Manon Pilaud[1][0009-0002-7586-6944] and Ian McCulloh[2][0000-0003-2916-3914]

[1] Johns Hopkins University, Baltimore, MD 21218, USA
mpilaud1@jhu.edu
[2] Johns Hopkins University, Baltimore, MD 21218, USA
imccull4@jhu.edu

**Abstract.** This study investigates potential indicators of coordinated influence activity among users participating in both r/Sino and r/China, two ideologically divergent Reddit communities focused on Chinese political discourse. Topic modeling and sentiment analysis are applied to all posts and comments authored by dual-subreddit users to construct a user–topic sentiment matrix. Individual sentiment patterns are compared to global topic baselines derived from the broader r/Sino and r/China populations. Behavioral profiling is performed using full user activity histories and metadata, incorporating measures such as lexical diversity, language consistency, account age, posting frequency, and karma distribution. Users exhibiting multiple behavioral anomalies are identified and examined within a subreddit co-participation network to assess structural overlap. The combined linguistic and behavioral analysis enables the identification of patterns consistent with inauthentic or strategically structured participation. These findings demonstrate the utility of integrating content and activity-based signals in the analysis of online influence behavior within contested information environments.

## 1 Introduction

Online platforms increasingly serve as arenas for geopolitical contestation, where narratives about states, societies, and conflicts are shaped, diffused, and strategically manipulated. Among these platforms, Reddit occupies a unique position due to its semi-anonymous structure and community-driven moderation model. Its decentralized architecture fosters ideological diversity, but also creates fertile ground for covert influence operations. This paper examines two prominent Reddit communities—r/Sino and r/China—which represent ideologically divergent spaces focused on Chinese political discourse. While r/China often features critical perspectives and Western-aligned discourse, r/Sino promotes narratives consistent with pro-government and nationalist positions, reinforced through selective moderation and content curation.

This ideological divergence provides a valuable setting to examine cross-community influence behavior. Users who actively participate in both communities may serve as narrative bridges—or, alternatively, as vectors for coordinated messaging efforts aimed at shaping or reframing contested discourse. Identifying the behaviors, sentiment patterns, and thematic alignments of these users offers insight into how influence operations may manifest in open-source environments. Such research aligns with emerging frameworks in social cybersecurity [1], which emphasize the need to detect and understand influence campaigns that exploit sociotechnical systems.

We build upon existing work in social network analysis and information operations detection [1]-[11], applying a combined methodology of topic modeling, sentiment analysis, and behavioral profiling to users active in both subreddits. By integrating content- and activity-based signals, our approach contributes to the growing literature on open-source intelligence (OSINT) methods for identifying coordinated online behavior in politically sensitive domains. By integrating linguistic and behavioral indicators, this research contributes to the identification of potential coordinated influence behavior and to methodological efforts for open-source monitoring of politically sensitive online spaces.

## 2    Background

Reddit has been the subject of increasing scholarly attention as a platform for ideological polarization, echo chambers, and strategic messaging [8]-[9],[12]. Much of the current literature has focused on U.S.-centric political topics or health misinformation [13]-[14], with relatively limited attention paid to Reddit's role in geopolitical influence operations. In contrast, platforms like Twitter and Facebook have been the primary focus of studies examining state-sponsored disinformation and bot behavior [7],[10],[15]-[16].

Yet Reddit presents unique methodological opportunities—and challenges—for influence detection. Its semi-anonymous architecture, decentralized moderation, and long-form discourse differentiate it from other social platforms and make traditional bot detection heuristics less reliable [8]-[9],[12]. Previous efforts to detect inauthentic activity on Reddit have employed indicators such as posting frequency, temporal regularity, lexical redundancy, and karma asymmetry [8]-[9],[12],[17]. However, most of these studies analyze users within isolated subreddits or narrowly defined thematic domains, without considering cross-community behavioral coherence.

Our work addresses this gap by examining discourse and behavior across ideologically opposed communities. We draw on social cybersecurity theory [1] to frame the potential for coordinated influence campaigns that operate beneath the surface of overt bot activity, using more nuanced strategies like tone modulation, narrative blending, and identity-driven engagement. This approach resonates with concepts from engagement manipulation and cognitive security articulated in military SNA contexts [9]-[11], where influence is operationalized not solely through automation, but through strategic manipulation of discourse networks and ecosystems.

We also position this work within the broader trajectory of computational propaganda research [18]-[19], which has highlighted the evolving nature of influence operations from crude spam campaigns to more sophisticated hybrid efforts involving both automated and human actors. Our focus on dual-subreddit users enables the detection of subtle alignment behaviors—users who may amplify specific narratives, maintain consistent sentiment profiles across topics, or exhibit stylistic irregularities that suggest non-organic participation.

By integrating topic and sentiment analysis with behavioral heuristics, we contribute a modular framework for identifying indicators of coordinated influence within Reddit. This study thus extends prior work in OSINT and influence detection by highlighting the importance of cross-community dynamics, affective profiling, and narrative evolution in the detection of strategic online messaging.

## 3 Methodology

Reddit posts and comments were collected from r/Sino and r/China using the Reddit API via PRAW and Selenium. The dataset includes post and comment-level metadata such as usernames, timestamps, subreddit affiliation, and full text content. In total, the corpus contains 999 posts and 6,036 comments from r/Sino and 930 posts and 7,641 comments from r/China. Users who posted or commented at least three times in each subreddit were retained, yielding a sample of 63 dual-subreddit participants.

All posts and comments authored by the 63 dual-subreddit users were preprocessed through lowercasing, tokenization, stopword removal, and lemmatization. Topic modeling was conducted using Gensim's implementation of Latent Dirichlet Allocation (LDA), with six topics trained on the tokenized dual-user corpus to uncover dominant discourse themes. To provide a comparative baseline, a second LDA model was trained on the full corpus of posts from both r/Sino and r/China, including users who were not active in both subreddits. This global model enabled validation of thematic alignment and supported downstream sentiment comparison.

Sentiment analysis was performed using the TextBlob polarity classifier. Sentiment scores were computed at the post level and aggregated by user and dominant topic to construct a user–topic sentiment matrix. Each user's topic-level sentiment was then compared to two baselines: (1) the average sentiment for that topic among all dual-subreddit users, and (2) the global average sentiment for that topic across all posts in the all-user model. This enabled identification of affective alignment or deviation relative to both peer and population-level discourse.

Full user activity histories were then aggregated to compute behavioral indicators. These included account age, total number of posts, karma distributions, lexical diversity, language detection, and posting irregularities. A set of heuristic-based behavioral flags was defined, and users exhibiting two or more flags were identified for further scrutiny.

To examine the structure of discourse environments, a subreddit co-participation network was constructed. Subreddits were linked based on shared users, and nodes associated with flagged users were visually distinguished. The final dataset was used to assess the extent of semantic alignment, affective consistency, and behavioral anomalies within and across ideologically polarized communities.

## 4 Findings

### 4.1 Cross-Community Participation and Activity Distribution.

Among 5,054 unique users who posted in r/China or r/Sino, 63 individuals (1.2%) met the criteria for dual participation, defined as posting or commenting at least three times in each subreddit. This cohort exhibited varied engagement levels, with a mean of 8.7 total posts or comments per user, a median of 5, and a high standard deviation of 11.7, indicating the presence of a small number of highly active contributors. On average, dual participants directed 44.1% of their activity toward r/China and 55.9% toward r/Sino, suggesting a slight engagement imbalance in favor of r/Sino. This distribution may reflect greater alignment with r/Sino's discourse norms or moderation policies. The high variance in activity levels also raises the possibility that a subset of users plays a disproportionate role in shaping cross-subreddit discourse.

### 4.2 Discourse Structure and Topic Alignment Across User Cohorts

To understand the narrative frames employed by users in r/Sino and r/China, Latent Dirichlet Allocation (LDA) was applied to two distinct corpora: one composed of posts and comments from users active in both subreddits ("dual users") and another drawn from the full population of participants in either community.

The dual-user model was trained on a tokenized corpus consisting of all posts and comments authored by the 63 users who met the minimum participation threshold in both communities. This model produced six topics that did not align cleanly with discrete rhetorical domains. Instead, the resulting topics reflected a diffuse and often blended discourse structure. The inter-topic distance map (Fig.1) revealed that only Topics 1 and 2 were positioned closely in semantic space, while the remaining topics were widely separated. This indicates that dual-subreddit users engaged across a broad and fragmented set of themes, with limited thematic clustering or rhetorical cohesion.

Topic 1, the largest topic (25.6% of tokens), seems to capture discourse around U.S.-China relations, particularly during the Trump era. It touches on trade issues, diplomatic policy, and broader geopolitical narratives. The dominance of terms like *china*, *trump*, *tariff*, and *american* suggests strong political and economic framing. Topic 2 (21.5% of tokens), appears to focus on China's role in the global economy, with key terms such as "china", "chinese", "world", "people", "would", "trade", "market", "money", "tariff", and "american" suggesting discussions related to international commerce, financial systems, and economic influence. Unlike Topic 1, which leaned more heavily into political discourse, this topic emphasizes economic interactions, including trade relations and market dynamics. The presence of terms like "rate", "restaurant", and "find" may indicate references to consumer markets or economic indicators, while words like "news" and "war" suggest ongoing coverage and potential geopolitical tensions influencing economic outcomes. Topic3, which accounts for 16.6% of tokens in the corpus, and appears to center around war, propaganda, and geopolitical discourse involving China and the U.S. Key terms such as "china", "war", "chinese", "american", "propaganda", "usa", "support", "taiwan", and "russia" suggest the topic covers military tensions, political narratives, and media framing related to national security and international conflict. The presence of emotionally charged and rhetorical terms like "never", "stop", "say", and "really" may indicate a polarized or strongly opinionated context. Topic 4, makes up 12.5% of tokens in the corpus. It appears to focus on political systems and ideological comparisons, especially in the context of China versus Western or European governance and values. Key terms include *china*, *chinese*, *european*, *economic*, *system*, *democracy*, *collapse*, *state*, and *europe*, indicating discourse around democracy vs. authoritarianism, economic models, and perceptions of societal decline or reform. Terms like *better*, *want*, *used*, *way*, and *think* suggest evaluative or comparative framing, potentially reflecting debates or opinions about governance and global influence. The language in topic 5 is more conversational and personal in tone, featuring frequent use of informal expressions like *dont*, *cant*, *well*, *lol*, and *going*. Core terms such as *chinese*, *china*, *american*, *trump*, *taiwan*, and *company* suggest the topic centers around individual or public opinion on geopolitical and economic issues, potentially reflecting everyday commentary on work, trade, politics, and identity. The topic is distinct from others due to its tone and positioning on the intertopic map, indicating a different type of discourse—less formal, more subjective or social-media driven. This topic is also smaller, containing 12.3% of tokens. Finally, the last topic (Topic 6), which accounts for 11.5% of tokens in the corpus. The topic appears to focus on cultural and ideological contrasts between China and the West, with core terms such as *china*, *western*, *chinese*, *american*, *taiwanese*, *social*, *government*, and *hypocrisy*. The presence of reflective or evaluative words like *know*, *see*, *think*, *need*, *power*, and *standard* suggests commentary or critique—possibly on norms, governance, and values. Terms like *food*, *live*, *world*, and *medium* hint at daily life or media as vehicles for expressing or examining these contrasts.

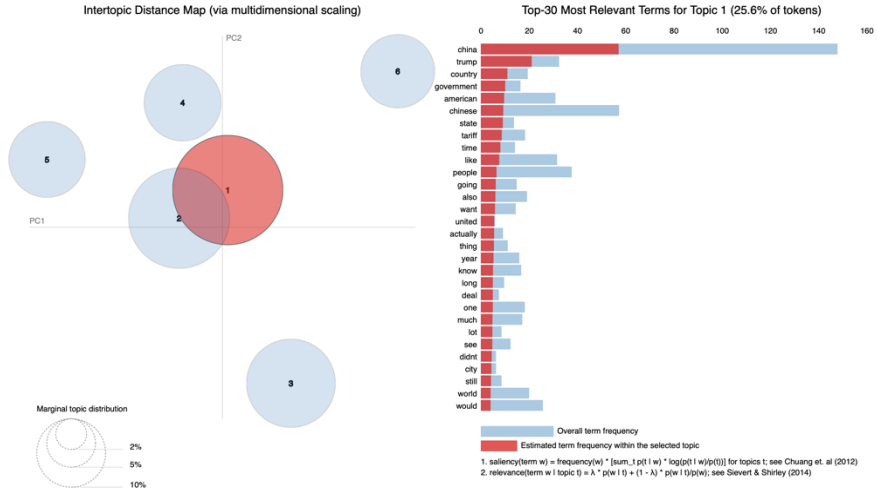

Figure 1. LDA topic model of dual users, active in both r/Sino and r/China

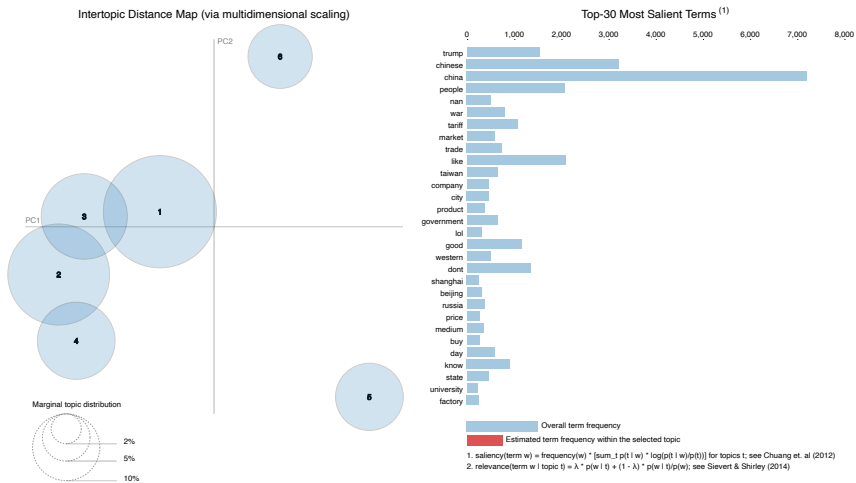

Figure 2. LDA topic model of all users, active in r/Sino or r/China

To establish a baseline for comparison, a second LDA model was trained on the full corpus of posts from r/Sino and r/China, including users who did not participate in both subreddits. Compared to the more diffuse and stylistically heterogeneous dual-user model, the global model exhibited greater thematic cohesion (Fig. 2). Four of its six topics clustered closely in the inter-topic distance map, indicating a more topically focused and centralized discourse among the general Reddit population. Topics were also more clearly defined, thematically distinct, and less emotionally varied. For example, discussions focused specifically on U.S.–China trade, manufacturing and tariffs, Taiwan-related diplomacy, study and travel logistics, and media propaganda. The language was generally neutral to formal, suggesting content drawn from news, policy commentary, or practical advice.

In contrast, the dual-user model presented a more fragmented and stylistically fluid discourse space. Topics overlapped heavily in both vocabulary and tone, with boundaries between themes such as politics, ideology, and identity less sharply drawn. Emotional expressions (*"lol"*, *"dont"*, *"shit"*, *"believe"*) appeared alongside policy terms (*"propaganda"*, *"democracy"*, *"war"*), indicating a blend of informal rhetoric and ideological signaling. The topic model surfaced broad, multi-thematic clusters, often mixing personal opinion with geopolitical references. This suggests that dual-users may operate across narrative boundaries, acting as bridges between ideological frames and diffusing narratives rather than reinforcing silos.

Crucially, while both models reflect shared concerns, such as U.S.–China rivalry, Taiwan, and political ideology, their structural presentation diverges sharply. The global model channels these themes into stable, self-contained topic zones, whereas the dual-user model embeds them in fluid, emotionally tinted discourse, potentially enhancing narrative flexibility or persuasive reach. This indicates that dual participants may not merely mirror broader Reddit discourse, but may reshape or redirect it, using their position across subreddits to modulate tone, reframe arguments, or inject ambiguity into polarized conversations.

### 4.3 Sentiment Divergence Across Topics

To assess affective alignment and deviation among dual-subreddit users, a user–topic sentiment matrix was constructed. Each Reddit post was assigned a sentiment polarity score using TextBlob, aggregated by user and dominant topic. Topic assignments were based on an LDA model trained on the full r/Sino and r/China corpus, allowing dual-user sentiment to be interpreted within the broader structure of Reddit discourse. User-level sentiment scores were then compared to global topic-level sentiment baselines derived from the same model. The result was a per-user sentiment profile, revealing both the tone and consistency with which individuals engaged across different themes.

Several notable patterns emerged. First, a subset of dual users expressed markedly more positive sentiment than the global average on topics that were otherwise neutral or negative. Most prominently, Topic 4, which focused on economic and trade discourse including terms like market, company, product, and tariff, had a global average sentiment of 0.094. However, among 7 dual users flagged as positive outliers, the average sentiment was 0.427, a sharp +0.333 deviation. This unusually elevated affect in a topic normally characterized by restrained or pragmatic tone may reflect an attempt to recast economic relations or market narratives in a more favorable light, possibly to project confidence or optimism around China's trade position.

Second, the analysis identified a group of low-variance users whose sentiment remained strikingly consistent across topics. These 5 users showed minimal deviation in tone, with an average sentiment of −0.011 and a standard deviation of just 0.029. Their posts were distributed across Topic 1 (broad discourse on China and its people), Topic 2 (U.S.–China geopolitical tensions), and Topic 3 (political ideologies and state structures). This uniformity of tone, regardless of the topic's typical emotional valence, could suggest an effort to maintain a controlled rhetorical stance, whether to appear neutral, mask intent, or maintain credibility across polarized environments.

Conversely, a larger group of dual users expressed significantly more negative sentiment than the general Reddit population on topics that were globally neutral or mildly positive. A total of 42 users were identified as negative sentiment outliers, with average sentiment of −0.167 in topics where the global average ranged from 0.054 to 0.119, resulting in downward deviations of up to −0.302. These users were most active in Topic 1 (China and culture), Topic 2 (geopolitical conflict, including Taiwan and trade), and Topic 4 (economic issues). The consistent presence of negative affect in these discourse zones suggests possible contrarian or disruptive messaging, potentially aimed at injecting cynicism or undermining prevailing narratives

Together, sentiment-based anomalies reinforce earlier structural findings: that dual-subreddit users do not merely reflect the tone and topic structure of the broader Reddit population. Instead, their affective profiles, whether unusually positive, persistently flat, or sharply negative, suggest deliberate engagement strategies, possibly intended to shape or modulate narrative dynamics in ideologically sensitive conversations.

**Table 1.** Dual-User LDA Topic Summary

| Topic | Top Keywords | Interpretation |
|---|---|---|
| **Topic 1** | "china","trump","country","govern ment","american","chinese","tariff", "state", "united", "deal" | US-China relations during the Trump era with economic emphasis. |
| **Topic 2** | "china", "chinese", "world", "people", "would", "trade", "market", "money", "tariff", "american" | Focuses on China's role in the global economy, highlighting trade, markets, tariffs, and financial influence. Emphasizes economic interactions and consumer dynamics over political discourse. |
| **Topic 3** | "china","war","chinese","american", "people","like","propaganda", "usa", "life","support","taiwan", "america", "russia", "russian", "current" | Focuses on geopolitical conflict and propaganda, especially relating to China, the U.S., and Taiwan. Highlights themes of war, media narratives, and international tensions. |
| **Topic 4** | "china", "chinese", "european", "economic", "system", "democracy", "collapse", "europe", "state", "used" | Centers on ideological and political comparisons, particularly between China and Western democracies, including perceived economic, government, and societal strengths or weaknesses. |
| **Topic 5** | "chinese", "china", "american", "people", "trump", "taiwan", "company", "buy", "work", "believe" | Captures informal, opinion-driven discussions involving China, the U.S., and related political economic themes. Conversational language suggests social media or everyday discourse tone. |
| **Topic 6** | "china", "western", "chinese", "american", "taiwanese", "social", "government", "hypocrisy", "power", "standard" | Explores cultural, ideological, and social contrasts between China, the West, and Taiwan. Includes themes of governance, societal values, and critique of power structures and perceived hypocrisy. |

**Table 2.** All-User LDA Topic Summary

| Topic | Top Keywords | Interpretation |
|-------|--------------|----------------|
| **Topic 1** | "china", "people", "dont", "like", "chinese", "know", "even", "really", "say", "care" | Captures broad, conversational discourse about China and its people, with informal tone and general sentiment. |
| **Topic 2** | "china", "trump", "war", "taiwan", "trade", "tariff", "military", "russia", "ukraine", "japan" | Focuses on U.S.–China geopolitical tensions, highlighting Trump, Taiwan, trade, and global security. |
| **Topic 3** | "china", "government", "western", "state", "democracy", "propaganda", "party", "communist", "freedom", "political" | Explores political ideologies and systems, contrasting China's government with Western democratic values. |
| **Topic 4** | "china", "market", "company", "product", "tariff", "manufacturing", "buy", "price", "industry", "export" | Focuses on manufacturing, trade, and economic products related to China and the global market. |
| **Topic 5** | "china", "city", "university", "student", "visit", "school", "visa", "foreigner", "shanghai", "beijing" | Captures study abroad, travel, and visa-related experiences, particularly in Chinese cities and schools. |
| **Topic 6** | "chinese", "video", "post", "news", "article", "propaganda", "reddit", "asian", "deleted", "watch" | Focuses on social media discourse, online content, and digital propaganda related to Chinese identity. |

### 4.4 Behavioral Anomaly Detection

In addition to linguistic and sentiment-based analysis, dual-subreddit users were evaluated for behavioral anomalies based on publicly available Reddit metadata. Each user was profiled using five criteria: account age relative to activity level, karma distribution, email verification status, lexical diversity, and detected language. These indicators were aggregated into a heuristic scoring framework, and users were assigned a flag count based on how many anomaly types they exhibited.

Among the 63 dual-subreddit users, 5 were flagged with two or more behavioral anomalies, suggesting elevated risk for inauthentic or coordinated activity. The most common anomaly was low lexical diversity, observed in 51 users, which may indicate templated language. Other notable flags included high activity paired with low link karma (5 users), comment–link karma imbalances (4 users), and non-English dominant language (1 user). Only 8 users exhibited no detectable anomalies by these measures.

In addition to these behavioral signals, 2 of the dual users were suspended by Reddit at the time of metadata collection. While account suspension is not definitive proof of inauthentic behavior, it further supports the need for scrutiny in cases where multiple behavioral and linguistic anomalies converge. This multi-indicator approach reinforces the observation that a small but meaningful subset of cross-community participants may be engaging in strategic, non-organic discourse patterns.

**Table 3.** Summary of Sentiment Outlier Patterns by Topic Among Dual-Subreddit Users

| Group | Topic | Num Users | Avg Sentiment | Global Avg Sentiment | Sentiment Deviation |
|-------|-------|-----------|---------------|----------------------|---------------------|
| Low Variance Users | 0 | 3 | -0.007 | 0.066 | -0.073 |
| Low Variance Users | 1 | 1 | -0.045 | 0.119 | -0.164 |
| Low Variance Users | 2 | 2 | 0.001 | 0.060 | -0.059 |
| Low Variance Users | 3 | 1 | 0.000 | 0.094 | -0.094 |
| Low Variance Users | 4 | 1 | -0.050 | 0.054 | -0.104 |
| Low Variance Users | 5 | 4 | 0.011 | 0.081 | -0.071 |
| Neg. Sent. Outliers | 0 | 7 | -0.236 | 0.066 | -0.302 |
| Neg. Sent. Outliers | 1 | 12 | -0.016 | 0.119 | -0.135 |
| Neg. Sent. Outliers | 2 | 5 | -0.190 | 0.060 | -0.250 |
| Neg. Sent. Outliers | 4 | 12 | -0.213 | 0.054 | -0.266 |
| Neg. Sent. Outliers | 5 | 6 | -0.167 | 0.081 | -0.249 |
| Pos. Sent. Outliers | 3 | 7 | 0.427 | 0.094 | 0.333 |

## 4.5    Network Structure and Subreddit Participation

The subreddit co-participation network of dual users reveals a densely clustered but strategically dispersed ecosystem, with users extending outward from *r/Sino* and *r/China* into a broad array of political, cultural, technological, and general-interest communities (Fig. 3). While subreddits like *r/AskChina, r/AskAChinese*, and *r/chinalife* serve as thematic extensions of China-related discourse, the broader network highlights a notable presence of flagged users across forums with strategic and global significance.

Flagged users are not limited to ideologically aligned or region-specific communities. Instead, they are embedded across structurally important and topically diverse subreddits, including *r/worldnews, r/technology, r/AskReddit*, and *r/europe*, as well as specialized forums tied to national security and economic narratives, such as *r/Huawei, r/StockMarket, r/Economics*, and *r/Futurology*. These subreddits are frequented by users interested in global finance, emerging technologies, and state–industry relations, present high-value targets for shaping public perception.

Rather than clustering at the network's edges, many flagged users occupy bridging positions that connect distinct topical clusters—serving as conduits between geopolitical, economic, and public-interest spaces. Frequent co-occurrence in triads like *r/worldnews–r/europe, r/AskChina–r/AskAChinese*, and *r/technology–r/AskReddit* indicates an ability to migrate narratives across domains, from tightly focused discussions of Chinese policy to high-traffic forums where geopolitical and economic discourse is mainstreamed.

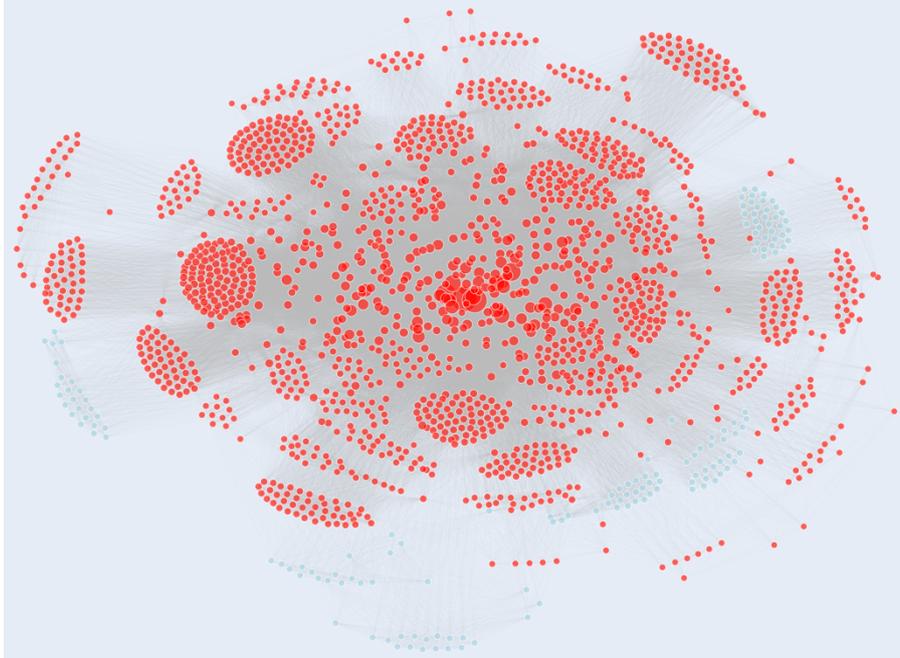

Figure 3. Network visualization of full sample of 1,819 subreddits. Subreddits with flagged user activity indicated in red.

Of the top 100 subreddits by degree, most show activity from flagged dual users. Across the full sample of 1,819 subreddits, 1,668 received activity from at least one flagged user. This expansive footprint, combined with earlier findings on sentiment anomalies and behavioral flattening, suggests that flagged users may act as narrative brokers or re-framers, subtly shifting the tone or framing of China-related content across multiple strategic domains. Their visibility in subreddits concerned with telecommunications infrastructure, financial systems, futures and global current events underscores a structural pattern that is not only broad but positionally optimized for influence.

## 5    Conclusion and Future Work

This study investigated user behavior across r/Sino and r/China—two ideologically opposed Reddit communities engaged in discourse about China—to surface indicators of coordinated inauthentic behavior. By focusing on users who participated in both communities, we identified a subset exhibiting linguistic, affective, and behavioral anomalies that deviate from the general Reddit population. These anomalies included topic-level sentiment shifts, unusually consistent or divergent affective profiles, and metadata patterns suggestive of non-organic engagement.

Our results indicate that dual-subreddit users may not simply act as ideological bridges, but could function as strategic amplifiers—modulating tone, reinforcing narratives, and diffusing discourse across communities. When considered in combination with subreddit co-participation patterns and engagement distribution, the findings suggest the possibility of engagement manipulation, where user behaviors (e.g., post timing, tone modulation, karma distribution) are leveraged to game platform dynamics and influence the visibility or perceived legitimacy of content. This aligns with broader concerns in social cybersecurity and computational propaganda research, which emphasize the evolving sophistication of influence campaigns that blend automated tools with coordinated human behavior.

The modular framework presented here—combining topic modeling, sentiment profiling, and heuristic-based behavioral analysis—offers a scalable approach for identifying suspicious patterns in open-source environments. It contributes both empirically and methodologically to the detection of subtle, persona-driven influence activity that may escape traditional bot-detection systems.

Future work should incorporate temporal dynamics, enabling the tracking of narrative evolution and user behavior over time to distinguish organic shifts from coordinated strategy. In addition, the flagged users identified in this study could serve as weak labels for developing supervised classifiers to detect coordinated inauthentic behavior more systematically. Integrating structural features (e.g., karma ratios), affective signatures (e.g., sentiment deviation from topic norms), and co-engagement networks could support the development of robust, generalizable detection pipelines.

As Reddit and similar platforms continue to shape geopolitical discourse, the ability to detect cross-community influence operations remains critical for platform integrity, public trust, and the broader aims of open-source intelligence.